# Bell Inequalities, Experimental Protocols and Contextuality.

Marian Kupczynski

Département de l'Informatique, Université du Québec en Outaouais (UQO), Case postale 1250, succursale Hull, Gatineau. Quebec, J8X 3X 7, Canada

**Abstract** In this paper we give additional arguments in favor of the point of view that the violation of Bell, CHSH and CH inequalities is not due to a mysterious non locality of Nature. We concentrate on an intimate relation between a protocol of a random experiment and a probabilistic model which is used to describe it. We discuss in a simple way differences between attributive joint probability distributions and generalized joint probability distributions of outcomes from distant experiments which depend on how the pairing of these outcomes is defined. We analyze in detail experimental protocols implied by local realistic and stochastic hidden variable models and show that they are incompatible with the protocols used in spin polarization correlation experiments (SPCE). We discuss also the meaning of "free will", differences between quantum and classical filters, contextuality of Kolmogorov models, contextuality of quantum theory (QT) and show how this contextuality has to be taken into account in probabilistic models trying to explain in an intuitive way the predictions of QT. The long range imperfect correlations between the clicks of distant detectors can be explained by partially preserved correlations between the signals created by a source. These correlations can only be preserved if the clicks are produced in a local and deterministic way depending on intrinsic parameters describing signals and measuring devices in the moment of the measurement. If an act of a measurement was irreducibly random they would be destroyed. It seems to indicate that QT may be in fact emerging from some underlying more detailed theory of physical phenomena. If this was a case then there is a chance to find in time series of experimental data some fine structures not predicted by QT. This would be a major discovery because it would not only prove that QT does not provide a complete description of individual physical systems but it would prove that it is not predictably complete.

## 1 Introduction

This year we celebrate 50 years of Bell's paper [1, 2]. In spite of a significant progress in understanding of the implications and the meaning of the violation of Bell, CHSH,CH and Eberhard inequalities (B-CHSH), in spin polarization correlation experiments (SPCE), there remains still a lot of confusion and diverging opinions. It can clearly be seen from the papers published in a special issue of Journal of Physics A devoted to '50 years of Bell's theorem' [3].

For many years a prevailing interpretation of the violation of B-CHSH in particular among the Quantum Information Community (QIC) has been a mysterious *quantum nonlocality* [2,4,5] and the papers of the authors who had different opinion were often ignored.

Analysing different proofs of B-CHSH several authors concluded [6-46], very often independently, that the violation of these inequalities invalidates only *counterfactual reasoning* or/and incorrect probabilistic models used to prove them . By no means the list of references given in this paper is complete.

This conclusion is supported by the fact that B-CHSH were found to be violated in the experiments performed in various domains of science in which the notion of locality was irrelevant [24,25,51-53] and even in classical mechanics [36]. In ingenious computer experiments Kristel Michielsen, Hans de Raedt and collaborators [47-50] reproduced event by event, in a local and causal way, the statistical distribution of outcomes of several quantum experiments including also SPCE.

Therefore the violation of B-CHSH neither proves the mysterious non locality of Nature nor the completeness of QT.

This opinion seems to be shared now by some scientists [54] , having close relations with QIC and/ or with experimental groups testing B-CHSH, who came also to the conclusion that *Bell locality* has nothing to do with *Einstein locality* and that the necessary assumption used in some probabilistic and in <u>all</u> non-probabilistic proofs of these inequalities is *counterfactual definiteness* (CFD) . We are very pleased to see it even if we did not find in [54] reference to several papers claiming the same and published many years ago.

CFD called also *realism* [55] , may be defined as :" This assumption allows one to assume the definiteness of the results of measurements, which were actually not performed on a given individual system. They are treated as unknown, but in principle defined values. This is in striking disagreement with quantum mechanics and the complementarity principle" [54].

Please note that the violation of the *realism*, used <u>in the restricted meaning</u> of CFD, does not mean that the World in which we are living has no real independent existence and that the Moon is not there when we are not looking at it .

As it was shown by Bohr [57] and Kochen and Specker [58] QT provides a contextual description of physical phenomena. This feature of QT is not reflected in local realistic hidden variables models (LRHVM) and in stochastic hidden variable models SHVM. This is why Theo Nieuwenhuizen summarized nicely questionable assumptions leading to B-CHSH as *the fatal (theoretical) contextuality loophole* [43].

Kolmogorov probabilistic models are contextual and Andrei Khrennikov constructed recently a non-signaling Kolmogorov probabilistic model of a realistic SPCE <u>with incorporated completely random choice of settings (i ,j)</u> in which CHSH cannot be proven [26].

It is forgotten sometimes that QT does not predict strict correlations in SPCE. The reason is that there are no sharp directions in Nature and QT predictions have to be treated as probability density functions and integrated over some small intervals [30.33.35].

If the outcomes of the measurements were created locally in *<u>irreducibly</u> random* way then even imperfect long range correlations violating B-CHSH would be impossible..

One may wonder why a picture of " *two random dices giving strictly correlated outcomes*" was not rejected as absurd from the very beginning. The reason for this was perhaps existence of different interpretations of QT and meanings, in the colloquial language, attached to the subtle notions of: probability, randomness, realism and *contextuality* .

In this paper in a simple language understandable by non- specialists we want to elucidate these and not only these issues and discuss in detail:

- various probabilistic models (Kolmogorov and hidden variable models) ,their differences and their intimate relation to experimental protocols
- strong correlations between the outcomes of far-away experiments which do not require any causal interaction between the distant experimental measuring devices
- the importance of time factor and data pairing in establishing the distant correlations
- notion of *contextuality* and its implications
- local contextual probabilistic models of SPCE , in which B-CHSH cannot be proven
- violation of B-CHSH as an argument against *irreducible quantum randomness*
- how can we test directly whether QT is predictably complete or is an emergent theory

## 2 Probabilistic Models and Experimental Protocols

2.1 Probability, Experimental Protocols and Bertrand Paradox

The intimate relationship between a probabilistic model and an experimental protocol became clear by the resolution of the Bertrand paradox [59, 60]. In 1889 Joseph Bertrand considered two concentric circles with radiuses R and r respectively and asked a question:" What is the probability P that a chord of a larger circle cuts a smaller circle if R=2r? Various equally justified geometrical proofs give P=1/2, P=1/3 or P=1/4. A solution to the paradox is simple each of these probabilities correspond to different random experiments, performed according to well defined protocols, which may be used to find the answer to Bertrand's question.

This is why Kolmogorov probabilistic models are contextual and to each random experiment is associated its own sample (probability) space Ω containing all different possible outcomes of the experiment and a corresponding probability distribution or more generally a probability measure.

2.2 Simple Classical Random Experiments

Let us consider a box containing 2 red and 1 black balls and two balls are drawn one after another.

- In the first experiment each draw is with replacement, a value of a random variable X= number of red balls drawn is registered and the experiment is repeated. X is taking three values 0, 1 or 2 therefore a sample space $\Omega=\{0,1,2\}$ and the probability distribution is given by: P (X=0)=1/9, P(X=1)=4/9, P(X=2)= 4/9.
- In the second experiment the draws are made without replacement therefore X is taking only two values 1 or 2, $\Omega=\{1,2\}$ and P(X=1)=2/3, P(X=2)=1/3.

If we have k red and m balls in a box and we draw n balls one after another with replacement after each draw then $\Omega=\{0,1,…n\}$ and the random variable X= number of red balls drawn obeys a standard binomial distribution B(n,p) where p= k/(k+m). If we draw without replacement X obeys a hypergeometric distribution.

## 2.3 Compatible Random Variables and "Attributive" Joint Probability Distributions

Let us now consider a box containing 4 balls: 2 red (one big and one small) and 2 black ( one big and one small) . Each ball has two attributes: color and size. With color we associate a random variable $X_1$ taking the values 1 for red and 0 for black. With size we associate a random variable $X_2$ taking the values 1 for big and 0 for small. The protocol of our experiment is the following: draw one ball, record the color and the size and after replacement repeat your experiment. Each outcome is a value of a bivariate random variable $X=(X_1, X_2)$, a sample space $\Omega=\{(1,1), (1,0),(0,1),(0,0)\}$ and a joint probability distribution is particularly simple: $P(X_1=i, X_2=j)=1/4$.

Please note that the ''measurement'' of the colour does not change the size, colour and size can be" measured" simultaneously or successively in any order. This is the assumption of *realism* according to which a measuring apparatus is registering preexisting and compatible properties of physical systems.

Having the data of this experiment we don't need to perform experiments in which we concentrate only on registering the colour or the size of the ball . For these restricted experiments the sample spaces are $\Omega_1=\{0,1\}$ and $\Omega_2=\{0,1\}$ and the corresponding probability distributions are simply marginal probability distributions $P(X_1=i)=P(X_1=i, X_2=1) + P(X_1=i, X_2=0)$ and $P(X_2=j)=P(X_1=1, X_2=j) +P(X_1=0, X_2=j)$ found using the joint probability distribution.

If we measure simultaneously the values of n random compatible variables $X_1 …X_n$ then we describe the experiment by a multivariate random variable $X=(X_1, X_2 … X_n)$, a common probability space $\Omega$, and "attributive" joint probability density function(AJPD) $p(x_1, …, x_n)$.

From $p(x_1, …, x_n)$ after summation or integration over (n − 1) variables we obtain n marginal probability density functions $p_{X_i}(x_i)$ describing different random experiments in which only one random variable $X_i$ is measured. Similarly after summation or integration over different (n-2) variables we obtain n(n-1)/2 marginal probability density functions $p_{X_i X_j}(x_i,x_j)$ describing different random experiments in which only a pair $(X_i, X_j)$ of random variables is measured etc.

## 3 Classical and Quantum Filters

### 3.1 Compatible Observables and Classical Filters

Macroscopic physical systems have well defined attributes (properties) $a_i$ i=1,2…n . Let us imagine that we have a mixed statistical ensemble (a beam) B of these systems and we want to choose only the systems having interesting for us attributes. For this purpose it is useful to have devices called filters.

<u>A classical filter $F_i$ or a macro selector is a device which passes the systems having an attribute $a_i$ and stops all the other. The filters operate according to Boolean Yes- or -No logic</u>

If we have n different attributes we have n filters corresponding to them. They can be put in any sequence chosen by us on the way of the beam B in function of our needs. The ensemble of these filters forms a classical lattice of filters having simple properties: $F_i F_i = F_i$, $F_i F_j = F_j F_i$
There exists also a maximal filter $F = F_1 F_2 \ldots F_n$ which transforms a mixed beam of the physical systems into a pure statistical ensemble in which all the systems have exactly the same properties.

The "attributive" joint probability distribution (AJPD) discussed in 2.3 describes a mixed statistical ensemble of physical systems characterized by n different compatible attributes.

3.2 Incompatible Observables and Quantum Filters

The formalism of QT was inspired by the optical experiments with polarized light.

Unpolarised light after passing through a linear (absorptive) polarizer, is transformed into linearly polarized light in the direction parallel to the axis of the polarizer and its (classically determined) intensity is reduced by half. Linearly polarized light passes without noticeable attenuation by a subsequent identical polarizer. The intensity of linearly polarised light after a passage through another polarizer is reduced according to Malus law: $I = I_0 \cos^2 \theta$ where $I_0$ is the initial intensity and θ is the angle between the light's initial polarization direction and the axis of the polarizer.

Discrete atomic spectral lines and photoelectric effect proved that the exchanges of energy between the electromagnetic field and the matter are quantized and the "carriers" of the quantized exchanged energy are *photons*. Thus we say now that an atom passes from a ground state to an excited state by absorbing a photon, that a light source is producing a beam of photons etc.

Linearly polarized monochromatic light is represented as a beam of linearly polarized photons each moving in the vacuum with the speed of light and carrying the energy $h\nu$. We cannot see photons, they are not point-like objects, and the only information we get is when a photon is absorbed by a very sophisticated photon detector which after several steps of signal enhancement produces a click what means : *a photon was detected.*

Coming back to the passage of the light represented as a beam of photons by a polarizer the intensity of the beam is now measured by counting the photons (clicks on the detectors) . Malus law is often rephrased by saying that each linearly polarized photon has a probability $p = \cos^2 \theta$ ,to pass through a polarizer, where θ is the angle between the photon's initial polarization direction and the axis of the polarizer.

As we saw in 2.1 a probability is a contextual property of a random experiment and not a property of a coin or of a photon.

<u>A quantum filter $F_i$ it is a device which creates a contextual property "i": "passing by $F_i$". A physical system having a property " i " have a probability $p_{ij}$ to pass by another filter $F_j$ acquiring</u>

after passage a new property " j " . Therefore the quantum filters are still idempotent $F_i F_i = F_i$ but in general they do not commute $F_i F_j \neq F_j F_i$ .

The lattice of quantum filters is isomorphic to the lattice of projectors on the subspaces of the Hilbert space. The quantum filters are not selectors of the pre-existing attributes of the physical systems but they are creators of contextual properties defined above.

Incompatible filters, as polarizers with non-parallel axes, create incompatible contextual properties which cannot be measured simultaneously and if measured in a sequence the previous contextual property is destroyed in a new measurement.

Therefore AJPD of these incompatible properties do not exist and different probabilistic models are needed in order to explain experiments in atomic physics. Such probabilistic models are constructed using QT.

In this paper we concentrate mainly on SPCE . Thus we will discuss a quantum model specific to these experiments. We try to avoid to talk about *pairs of photons* and we talk only when it is possible about signals produced by a source and clicks produced by the detectors.

## 4 Generalized Joint Probability Distributions

4.1 SPCE and coincidence loophole

SPCE can be described as below:

- A pulse from a laser hitting a non- linear crystal produces two correlated signals propagating in opposite directions.

- When we place polarizing beam splitters (PBS) in front of distant detectors we obtain two time series of clicks which are correlated.

- In order to eliminate possible causal influences between distant experimental set-ups the settings of PBS can be chosen in a systematic or in a random way when the signals were already produced by the source but they have not yet arrived to distant laboratories.

In idealized model of SPCE a source is sending two signals ("pairs of photons") which should arrive to distant detectors at the same time and produce the coincident clicks on some of them. The experimental situation is much more complicated since the clicks are not registered at the same time and one has to decide which clicks are correlated by introducing specific time windows and deciding how to use them in order to define coincident clicks.

Already in 1986 Pascazio [61] pointed out some pitfalls inherent in the coincidence mechanism. In 2004 Larsson and Gill [62] demonstrated that the coincidence determination can have a detrimental effect on Bell tests. These experimental problems are called sometimes *coincidence loophole*. An instructive discussion of different methods used to establish coincidences in SPCE may be found in a recent paper [63] .

4.2 Data Pairing and Generalised Joint Probability Distributions

Any classical or quantum probabilistic model in order to make predictions for the correlations observed in SPCE has to introduce a generalised joint probability distribution (GJPD) of outcomes of distant experiments.

GJPD is compared with empirical joint frequency distribution determined from two time series of data obtained in distant laboratories. To find these empirical joint probability distributions one has to adopt a criterion how distant observations are paired and the outcome may depend on the criterion used.

The problem of pairing of data coming from different experiments is more general than coincidence pairing in SPCE. Let us consider two experiments x and y in which we register two time series of experimental outcomes $S_1=\{a_1,a_2,... a_n ...\}$ and $S_2=\{b_1,b_2,..b_n ...\}$. In general using descriptive statistics we can find corresponding sample means, sample standard deviations and the empirical frequency distributions or histograms. In order to study correlations between these two sets of data we have to define a pairing of observed outcomes. For example we may create samples $S_{1k}=\{ (a_1, b_k), (a_2, b_{k+1}), (a_3, b_{k+2})...\}$ or a sample $S_R$ containing pairs $(a_s, b_t)$ chosen at random from $S_1$ and $S_2$ respectively.

If the experiments x and y consist on measuring some independent random variables A and B then $S_1$ and $S_2$ are simple random samples drawn from some distributions $p_A(a)$ and $p_B(b)$. In this case $S_{1k}$ and $S_R$ are some samples drawn from GJPD $p_{AB}(a,b) = p_A(a)p_B(b)$ and there is no correlation between the outcomes of x and y.

In general different pairings produce samples from different GJPDs since the only constraint we have is that $p_A(a)$ and $p_B(b)$ should be marginal probability distribution from $p_{AB}(a,b)$. Various problems related to non- uniqueness of GJPDs were discussed recently in different context and using different terminology by Dzhafarov and Kujala [51-53].

4.3 Alice, Bob, Charlie and Long Distance Correlations

Let us consider a simple example.

Imagine that Charlie using fair coin tosses or another randomizer creates a sequence $S_1=\{1001110100...\}$. Next using $S_1$ he creates a second sequence $S_2=\{0110001011...\}$ by replacing 1 by 0 and vice versa. He sends $S_1$ to Alice's computer and $S_2$ to Bob's computer. The distant computers output 1 when they receive 1 and -1 when they receive 0. If the pairing $S_{11}$ is used we get strong anti-correlations E (AB) = -1 but if the pairing $S_R$ is used E (AB) = 0, $p_{AB}(a,b) = p_A(a)p_B(b) = 1/4$ and outcomes are uncorrelated.
Similarly any strictly increasing time series of outcomes obtained in completely unrelated experiments x and y are strongly correlated if $S_{1k}$ pairings are used and become uncorrelated if the pairing $S_R$ is used.

To conclude: the type of correlation may depend strongly on how distant observations are paired and  <u>strong long distance correlations do not require any causal interactions</u> or communications <u>between Alice's and Bob's experimental set-ups</u>.

# 5 Quantum Contextual Probabilistic Model for SPCE

5.1 Operators and density matrices

In QT signals prepared by a source in SPCE are described by density matrices ρ and physical observables measured (random variables *A* and *B* describing the outcomes on the distant detectors) by hermitian operators $\hat{A}_1 = \hat{A} \otimes I$ and $\hat{B}_1 = I \otimes \hat{B}$. The correlations between outcomes observed by Alice and Bob for each fixed setting of PBS are found using conditional covariance between *A* and *B* [37,38]:

$$\text{cov}(A, B | \rho) = E(AB | \rho) - E(A | \rho) E(B | \rho) \tag{1}$$

where $E(A|\rho) = Tr\rho\hat{A}_1$, $E(B|\rho) = Tr\rho A\hat{B}_1$ and $E(AB|\rho) = Tr\rho\hat{A}_1\hat{B}_1$.

The marginal expectation values do not depend on distant experimental settings thus QT respects *Einstein locality=non-signalling*. Specific GJPD for *A* and *B*, predicted by QT, corresponds to coincidence pairing of the clicks on distant detectors. The model is contextual because the triplet $\{\rho, \hat{A}_1, \hat{B}_1\}$ changes if a system preparation or PBS settings change. For each setting QT provides a specific Kolmogorov model.

5.1 CHSH Inequalities

When a mixed quantum ensemble is prepared by a source:

$$\rho = \sum_{i=1}^{k} p_i \rho_i \otimes \tilde{\rho}_i \tag{2}$$

then

$$E(AB | \rho) = \sum_{i=1}^{k} p_i E(A | \rho_i) E(B | \tilde{\rho}_i) \tag{3}$$

and if $|E(A|\rho_i)| \leq 1$ and $|E(B|\tilde{\rho}_i)| \leq 1$ we obtain CHSH inequalities which were first derived using various hidden variable models:

$$| E(AB | \rho) - E(AB' | \rho) | + | E(A'B | \rho) + E(A'B' | \rho) | \leq 2 \tag{4}$$

where A, A', B and B' correspond to Alice's and Bob's different settings.

If ρ is given by (2) all predictions of QT can be reproduced by some SHVM discussed in the next section.

The signals produced in SPCE are not described by (2) and if one neglects background radiation and all experimental imperfections then according to QT a source is preparing "photon pairs" in perfectly entangled spin singlet state : $\rho = |\Psi\rangle\langle\Psi|$ and (4) may not be proven.

# 6 Hidden Variable Probabilistic Models for SPCE

## 6.1 LRHVM or *Bertlmann's Socks Model*.

In this model a source is producing a statistical classical mixed ensemble of "pairs of photons" labelled by $\lambda = (\lambda_1, \lambda_2)$ having well defined spin projections in all directions. After passing by a given PBS a photon is registered by one of the detectors what is interpreted as a reading of the pre-existing spin projection of this particular photon on the axis of PBS equal to $\pm 1$. Alice's and Bob's experimental outcomes in a PBS setting (i,j) are the values of two functions $A_i((\lambda_1)$ and $B_j(\lambda_2)$. The simplest probabilistic LRHV model is given by:

$$E(A_i B_j) = \sum_{\lambda \in \Lambda} P(\lambda_1, \lambda_2) A_i(\lambda_1) B_j(\lambda_2) \tag{5}$$

where $\Lambda$ is a set of all labels and $P(\lambda_1, \lambda_2)$ describes the probability distribution of them.

Please note that (5) does not define a Kolmogorov model because $(\lambda_1, \lambda_2)$ it is not an outcome of a random experiment but a label of a pair. However the model (5) is isomorphic to Kolmogorov model using AJPD.

Let us see what happens if we use only two different experimental settings on each side. For each "pair" labelled by $(\lambda_1, \lambda_2)$ we have 8 possible outcomes thus we can replace $(\lambda_1, \lambda_2)$ by $\omega = (a_1, a_2, b_1, b_2) \in \Omega$ where $a_i = \pm 1$ and $b_j = \pm 1$ and $P(\lambda_1, \lambda_2)$ by AJPD: $P(\omega) = P(a_1, a_2, b_1, b_2)$. Using this isomorphism we can rewrite (5) for i=1 and j=2 as:

$$E(A_1 B_2) = \sum_{\omega \in \Omega} a_1 b_2 P(a_1, a_2, b_1, b_2) = \sum_{\omega \in \Omega_{12}} a_1 b_2 P_{A_1 B_2}(a_1, b_2) \tag{6}$$

where $\Omega_{12} = \{\text{all } (a_1, b_2)\}$, $P_{A_1 B_2}(a_1, b_2)$ is a standard marginal distribution obtained from $P((a_1, a_2, b_1, b_2)$ by summing over $a_2$ and $b_1$. Please note that sample space $\Omega$ contains exactly 8 elements and $\Omega_{12}$ exactly 4 elements.

In this model "pairs of photons" are treated like pairs of socks which can be, for example, large or small, red or black. One sock from a pair is sent to Alice and another to Bob and different settings correspond to registering size or colour. This is why the model (5) was called by Bell: *Bertelsmann's Socks Model*.

<u>The experimental protocol consistent with (6) is similar to the protocol discussed in 2. 3 : take the first "pair of photons" measure $(a_1, a_2, b_1, b_2)$ and take another pair. It is obvious that this protocol cannot be implemented and has nothing to do with the protocol used in SPCE.</u>

The model (6) is clearly incompatible with QT and with SPCE since it assumes the existence of <u>the non-vanishing</u> joint probability distribution of the incompatible physical observables.

The models (5) and (6) can be generalised by allowing non registration of some clicks. In this case $a_i = \pm 1$ or 0 and $b_j = \pm 1$ or 0, $\Omega$ contains 81 elements and $P((a_1, a_2, b_1, b_2)$ exists and is well defined.

In some sense the existence of AJPD is closely related to CFD discussed in the introduction.

6.2 SHVM or *Pairs of Dices Model*.

We will discuss now a probabilistic meaning of SHVM [2, 5, 64 ]:

$$P(a,b|x,y) = \sum_{\lambda \in \Lambda} P(\lambda_1, \lambda_2) P(a|x, \lambda_1) P(b|y, \lambda_2) \tag{7}$$

From (7) we obtain immediately:

$$E(AB|x,y) = \sum_{\lambda \in \Lambda} P(\lambda_1, \lambda_2) E(A|x, \lambda_1) E(B|y, \lambda_2) \tag{8}$$

In model (7) we have two families of <u>independent</u> random experiments one on Alice's side labelled by ( x, $\lambda_1$) in which outcomes *a* are produced and another on Bob's side labelled by ( y, $\lambda_2$) in which outcomes *b* are produced . The corresponding probability distributions are $P(a|x, \lambda_1)$ and $P(b|y, \lambda_2)$ respectively. Labels ($\lambda_1$, $\lambda_2$) are generated by a computer according to a probability distribution P ($\lambda_1$, $\lambda_2$) and for each couple of distant experiments the corresponding GJPD is a product $P(a|x, \lambda_1) P(b|y, \lambda_2)$. This factorization is called *Bell locality* and it has nothing to do with *Einstein locality*.

According to (8) E (AB| x, y) are estimated according to the following protocol:

1. Generate a label ($\lambda_1$, $\lambda_2$)
2. Perform independent random experiments ( x, $\lambda_1$ ) and (y, $\lambda_2$) many times
3. Find estimates $\hat{E}(A|x, \lambda_1)$ and $\hat{E}(B|y, \lambda_2)$
4. Go to 1
5. After N loops find an estimate $\hat{E}(AB|x, y)$ as the average of $\hat{E}(A|x, \lambda_1) \hat{E}(B|y, \lambda_2)$

As we see this protocol cannot be implemented in SPCE. We cannot repeat the experiments on the same "pair of photons".

If the <u>same</u> hidden parameter space $\Lambda$ and P ($\lambda$) are used for different settings (x, y) and if all $|E(A|x, \lambda_1)| \leq 1$ and $|E(B|y, \lambda_2)| \leq 1$ then one can prove CHSH inequalities (4).

A more detailed discussion of the SHVM and its relation to the assumption of *irreducible randomness* of quantum measurements may be found in [36-38].

In SHVM "pairs of photons" are treated as *pairs of random dices*. Charlie prepares a particular set of correlated random dices and sends one dice to Alice and another dice to Bob. They are rolling their dices (repeating the experiments (x,y)) several times and recording each time their outcomes . Charlie sends another pair etc. Some non- vanishing correlations may be created in this way [37].

The model (7) describes (x,y) as a mixed ensemble of pairs of independent random experiments and its violation has no implication on the validity of *Einstein locality*.

6.3 Contextual Probabilistic Local Models for SPCE

In QT to each fixed setting (x,y) is associated its own Kolmogorov sample space $\Omega_{xy}$ and GJPD P(a,b|x,y) predicted by the theory. As we discussed in 3.2 and 6.1 P(a,b|x,y) cannot be obtained as marginal distribution from some non–vanishing GJPD of outcomes from all possible settings.

Contextual local probabilistic models (CLPM) for SPCE may be defined as follows :

$$E(AB | x, y) = \sum_{\lambda \in \Lambda_{xy}} P(\lambda_1, \lambda_2) P_x(\lambda_x) P_y(\lambda_y) A(\lambda_1, \lambda_x) B(\lambda_2, \lambda_y) \tag{9}$$

where $A(\lambda_1, \lambda_x)$ and $B(\lambda_2, \lambda_y)$ are equal $\pm 1$ or 0. In each fixed setting (x,y) we have 2 detectors on both sides. The parameters $(\lambda_1, \lambda_2)$ describe the signals correlated by the source how they are perceived by the respective polarizers and $(\lambda_x, \lambda_y)$ are the intrinsic parameters of polarizers.

The model (9) is not a Kolmogorov probabilistic model but it is a model defining a specific detailed invisible internal protocol of a random experiment consistent with the idealized experimental protocol of SPCE.

In the moment of measurement invisible intrinsic parameters $\lambda=(\lambda_1, \lambda_2, \lambda_x, \lambda_y)$ determine locally which of 9 possible experimental outcomes are produced on both sides of the experiment in a chosen time window : two clicks on each side, a click on only one of the detectors and absence of a click. We are talking here about idealized SPCE in which the time window is such that there is no more than one click observed on each of the sides.

If one is not interested in details of the internal protocol then the probabilistic model (9) can be written in a compact notation for an experimental setting (i,j):

$$E(AB | i, j) = \sum_{\lambda \in \Lambda ij} P_{ij}(\lambda_i, \lambda_j) A(\lambda_i) B(\lambda_j) . \tag{10}$$

When written in this form the model (9) is isomorphic to a Kolmogorov model:

$$E(A_i B_j) = \sum_{\omega \in \Omega_{ij}} a_i b_j P_{A_i B_j}(a_i, b_j) \tag{11}$$

Now, in contrast to (6) $P_{A_i B_j}(a_i, b_j)$ are not marginal probabilities from some GJPD on a common probability space $\Omega$ because such GJPD does not exist. We added also the outcomes 0 on both sides thus each $\Omega_{ij}$ contains now 9 elements instead of 4 . Of course using (11) one may reproduce quantum predictions for any experimental setting (i, j) . It is obvious that (11) does not allow to prove Bell, CHSH and other inequalities.

6.4 Particular Features of Models Defining Probabilistic Internal Protocols .

As we saw in 2.3 if we perform a summation or an integration of a joint probability distribution in a Kolmogorov model we obtain various marginal probability distributions describing **feasible** experiments.

It is not a case if we perform a summation over $(\lambda_x, \lambda_y)$ in (9):

$$E(AB | x, y) = \sum_{\lambda_1, \lambda_2} P(\lambda_1, \lambda_2) \bar{A}(\lambda_1) \bar{B}(\lambda_2) \qquad (12)$$

where

$$\bar{A}(\lambda_1) = \sum_{\lambda_x} P_x(\lambda_x) A(\lambda_1, \lambda_x) \qquad (13)$$

and $\quad \bar{B}(\lambda_2) = \sum_{\lambda_y} P_y(\lambda_y) A(\lambda_2, \lambda_y)$ .

From a point of view of mathematics the equations (9) and (12-13) are equivalent and give the same numerical outcomes. However (12) and (13) define completely different internal protocols which cannot be implemented in SPCE:

1. Generate a first pair of ($\lambda_1, \lambda_2$).
2. Generate k pairs of ($\lambda_x, \lambda_y$).
3. Calculate k values of $A(\lambda_1, \lambda_x)$ and average them to find an estimate of $\bar{A}(\lambda_1)$.
4. Calculate k values of $B(\lambda_2, \lambda_y)$ and average them to find an estimate of $\bar{B}(\lambda_2)$.
5. Multiply the estimates obtained in the points 3 and 4 and output them.
6. Go to point 1 and loop N times
7. Output the average of N outputs found in the point 5 as an estimate of E(A,B|x,y).

As we showed above the summation or integration over some variables in a probabilistic model with invisible intrinsic parameters leads in general to a new probabilistic model describing a random experiment using different experimental protocol which cannot be implemented.

Thus one is not allowed to replace the model (9) by the model (12, 13) and one cannot prove CHSH [35-38].

The model is no signalling since $E(A | x, y) = E(A | x) = \sum_{\lambda_1, \lambda_2, \lambda_x} P(\lambda_1, \lambda_2) P_x(\lambda_x) A(\lambda_1, \lambda_x)$ .

We studied recently with Hans De Raedt, in a completely different context, some random experiments operating according to intrinsic probabilistic protocols similar to (9) and (12-13). We discovered dramatic differences between huge finite samples generated by Monte Carlo simulations when different internal probabilistic protocols were used [65].

The description of random experiments in terms of internal experimental protocols is different and more detailed than the description in terms of standard probabilistic models.

## 7 Non-Kolmogorovness and Contextuality versus Realism

7.1 Khrennikov Model for SPCE with Randomly Chosen Settings

Any random experiment in which the frequency distributions of different possible outcomes stabilize can be described by a Kolmogorov model and a specific sample space $\Omega$. This is why *nonKolmogorovness* used in [6-8,22] is now called *contextuality*.

Khrennikov constructed a Kolmogorov model [26] for SPCE with randomly switched settings (i, j). On each side of the experiment we have now two PBS and two pairs of detectors and after switching the signals are sent to a different pair of PBS. We may record each outcome of the experiment as a vector $\omega=(i,j, \omega_{11}, \omega_{12}, \omega_{21}, \omega_{22}, \omega'_{11}, \omega'_{12}, \omega'_{21}, \omega'_{22})$ where i,j= 1or 2 and $\omega_{ks}, \omega'_{mt}=$ 0 or 1. $\Omega$ is a subset of a set of $2^{10}$ vectors such that only 4 components are different from zero.

Assuming that there is no background radiation and that exactly one of the detectors on each side produces a click Khrennikov introduces specific random variables on $\Omega$ : $A_i$ and $B_j$ taking three values ±1 or 0. They are chosen in such a way that $E(A_i, B_j|i,j)$ correspond to the experimentally observed correlations to be compared with quantum predictions.

There exists on $\Omega$ a non-vanishing GJPD consistent with time coincidence pairing of distant outcomes and with non-signaling condition (*Einstein locality*). For any pair of random variables $(A_i, B_j)$ there exist also non- vanishing GJPDs. These non-vanishing GJPDs for $(A_i, B_j)$ cannot be obtained as marginal distributions from the <u>vanishing joint probability distribution</u> for $(A_1, A_2, B_1, B_2)$ in contrast to the models with attributive joint probability distributions discussed in 2.3. As Khrennikov shows for conditional expectation values $E(A_i, B_j|i,j)$ one may not prove CHSH .

7.2 Free Will and *Contextuality* of Quantum Mechanics.

A possibility of choosing randomly experimental setting is called *free will* or *freedom of choice* and is formulated often as in [63]: "The measurement setting distribution does not depend on the hidden variable, or equivalently, the probability measure P does not depend on the measurement settings".

This statement is incorrect. Of course settings can be chosen in an arbitrary way, random or not, and a choice does not dependent on intrinsic variables used in a model. Nevertheless the probability measures in (9-11) and in QT depend on the settings chosen because of *contextuality* of QT. The setting (i, j) it is not a couple of numbers it is a pair of polarizers described by their own intrinsic variables. Many proofs using causal incomplete directed acyclic graph (DAG) [55] miss this point.

Bohr strongly insisted on the wholeness of quantum experiment and on the important <u>active</u> role played by the measuring instruments. Using incompatible experimental settings we obtain only a knowledge about *complementary* properties of the physical systems. Kochen and Specker [58] gave a nice proof of contextual character of quantum observables.

We saw in 3.2 that quantum filters are not selectors of the pre-existing attributes but they are creators of new properties. A very detailed discussion of quantum measurements was given recently in [66].

Let us resume this short subsection by a citation from our old paper [30] :"a value of a physical observable associated with a pure quantum ensemble and in this way with an individual physical system, being its member, is not an attribute of the system revealed be a measuring apparatus; it turns out to be a characteristic of this ensemble created by its interaction with the measuring device. In other words the QM is a contextual theory in which the values of the observables assigned to a physical system have only meaning in a context of a particular physical experiment".

7.3 Counterfactual Definiteness and Realism

The counterfactual reasoning is also used in all the proofs in which finite experimental samples of size N were studied and Bell and CHSH inequalities were proven by using the large N limit [55,67-70].

The problem of the impact of finite statistics on the conclusions of the experimental tests of various inequalities was analyzed carefully by Richard Gill [55, 67, 70] who derived probabilistic bounds on the violation of CHSH in finite samples and came to the conclusion that perhaps decisive tests of these inequalities were impossible.

For many years he has been a virulent defender of Bell Theorems and of *quantum nonlocality* [67, 70]. He realized recently that the essential assumption in his finite sample proofs of CHSH was CFD and not *locality*. However he still claims that: "we can keep quantum mechanics, locality and freedom… taking quantum randomness very seriously: it becomes an irreducible feature of the physical world, a "primitive notion"; it is not "merely" an emergent feature" [55].

This statement is simply incorrect since as we discussed in detail in [38] the *irreducible randomness* of quantum measurements is able to produce only the same type of correlations as SHVM which as we showed in 6.2 satisfy CHSH . On the contrary as we saw in subsection 6.3 and in [38] a significant violation of CHSH would rather give arguments in favour of *local determinism* underlying quantum measurements and not in favour of *irreducible randomness*.

Using CFD one may only generate finite samples consistent with LRHVM or SHVM this is why in the limit when N tends to infinity they satisfy CHSH and for a finite N ,as Gill proved, the probability of observing a significant violation is small. Finite samples studied in [55, 67-70, 71] are all of this type and they do not correspond to finite samples created in SPCE.

Because of *contextuality* a counterfactual N x 4 spreadsheet discussed in [55] has nothing to do with observed experimental outcomes. Experimental outcomes are not predetermined but created in the moment of measurements. If we have four samples of size N obtained in each of four possible settings (A,B), (A,B'), (A',B) and (A',B') we have four different spreadsheets which cannot be deduced from N x 4 or (4N) x 4 spreadsheets discussed in [55].

## 8 Statistical Contextual Interpretation of QT.

In statistical and contextual interpretation of QT [24, 25, 30, 34, 35, 72] :

- A wave function $\Psi$ is associated with a pure ensemble of physical states. A hermitian or self-adjoint operator $\hat{O}$ is associated with a physical observable O. A couple ($\Psi$ , $\hat{O}$ ) is used to find the probability distribution of outcomes of measurements of the observable O performed on this ensemble in a particular experimental context.

- Experimental outcomes are not predetermined but created in the moment of the measurement (*contextuality*).

- A wave packet reduction is neither instantaneous nor non–local and a reduced state vector describes only a preparation of another ensemble of physical states having particular properties.

- In particular for EPR type experiments ,such as SPCE, a state vector describing a system II obtained by the reduction of the entangled state vector of two physical systems I+II describes only the sub-ensemble of the systems II being the partners of those systems I for which the measurement of some observable gave the same specific outcome.

If a state vector was treated **incorrectly** as <u>an attribute</u> of an individual physical system which can be modified instantaneously by a measurement performed in a distant location   a *spooky action at a distance* seemed to be unavoidable.

If statistical interpretation is adopted one may investigate  whether QT provides a complete description of  individual physical systems or even whether QT provides a complete description of the experimental data. These questions cannot be answered by proving  no-go theorems or by philosophical arguments. They can only be answered by a detailed study of  experimental time series of data.

## 9  Predictable Completeness of QT and How to Test It.

If  there exists  more detailed sub-quantum description of physical phenomena then:

   (1) Each pure quantum statistical ensemble is a mixed statistical ensemble

   (2) It is possible that there is more information in the experimental data than predicted by QT.

Many years ago in a different context we studied the differences between pure and mixed statistical ensembles [73] and we indicated the tests called by us *purity test* which might be used to check whether the ensemble of the experimental outcomes is "pure" or not.

The idea is simple in a pure statistical ensemble all sub-ensembles have the same properties as the initial ensemble. If the ensemble is mixed then by studying its sub-ensembles we may discover, in principle, significant differences.

Initially we wanted to apply these tests to check whether some high energy scattering experiments should be described by pure or mixed quantum states [74,75] but we also noticed their importance for a more general problem of completeness of QT [28,29]

The aim of most of physical experiments is to compare empirical probability distributions with quantum probabilistic predictions. Therefore all fine structures of time series of data if they existed would be averaged out and not discovered.

In order to discover such fine structures in time series of data *purity tests* are not sufficient and one has to use more detailed tests invented by statisticians for this purpose [65,76-78].

## 10  Conclusions.

We discussed in detail an intimate relation between a probabilistic model and a corresponding experimental protocol and explained why experimental protocols defined   by local realistic hidden  and stochastic hidden variable models are inconsistent with the experimental protocols used in SPCE.  These models suffer from so called "*contextuality loophole*" and do not include in a proper way intrinsic parameters describing measuring devices in the moment of measurement.

We also pointed out that correlations do not require any causal interaction between distant experimental set-ups. We discussed the differences between attributive joint probability distributions and the generalized joint probability distributions for the outcomes of distant experiments and their dependence on how the pairing of distant outcomes is done.

Finally we proposed a contextual probabilistic model in which none of the famous inequalities could be proven but in which the existence of long distance imperfect correlations in SPCE could be intuitively explained.

We concluded that the violation of various inequalities does not prove the breakdown of *Einstein locality* and it only invalidates some assumptions used to prove these inequalities in particular CFD. This point of view seems to gain recently a large consensus <u>as it should</u>. In our opinion *non-locality* appearing in some sub-quantum descriptions of quantum phenomena e.g. Bohmian mechanics is of different origin and it does not prove that *Einstein locality* should be abandoned.

Strong correlations observed in SPCE would not be possible if the outcomes were produced in *irreducibly random* way since all the correlations created between signals at a source would be destroyed. Therefore the violation of different inequalities seems to give an argument in favour that QT perhaps might be emergent from some underlying local and deterministic theory.

The question whether QT is emergent or predictably complete can be, in principle, answered by more detailed study of experimental time series of data and we hope that experimentalists will one day understand its importance.

**Acknowledgements** I am grateful to UQO for a travel grant and to Andrei Khrennikov for his hospitality and for the invitation to give a talk during this interesting QTPA conference.